\documentclass[reprint,superscriptaddress,aip,apl]{revtex4-2}
\usepackage{lineno}
\usepackage[separate-uncertainty=true]{siunitx}
\usepackage{graphicx}

\begin{document}

\title{Measurements of a quantum bulk acoustic resonator using a superconducting qubit}

\author{M.-H. Chou}
\affiliation{Pritzker School of Molecular Engineering, University of Chicago, Chicago IL 60637, USA}
\affiliation{Department of Physics, University of Chicago, Chicago IL 60637, USA}
\author{\'E. Dumur}
\altaffiliation[Present address: ]{Universit\'e Grenoble Alpes, CEA, INAC-Pheliqs, 38000 Grenoble, France}
\affiliation{Pritzker School of Molecular Engineering, University of Chicago, Chicago IL 60637, USA}
\affiliation{Argonne National Laboratory, Argonne IL 60439, USA}
\author{Y. P. Zhong}
\affiliation{Pritzker School of Molecular Engineering, University of Chicago, Chicago IL 60637, USA}
\author{G. A. Peairs}
\affiliation{Department of Physics, University of California, Santa Barbara CA 93106, USA}
\affiliation{Pritzker School of Molecular Engineering, University of Chicago, Chicago IL 60637, USA}
\author{A. Bienfait}
\altaffiliation[Present address: ]{Universit\'e de Lyon, ENS de Lyon, Universit\'e Claude Bernard, CNRS, Laboratoire de Physique, F-69342 Lyon, France}
\affiliation{Pritzker School of Molecular Engineering, University of Chicago, Chicago IL 60637, USA}
\author{H.-S. Chang}
\affiliation{Pritzker School of Molecular Engineering, University of Chicago, Chicago IL 60637, USA}
\author{C. R. Conner}
\affiliation{Pritzker School of Molecular Engineering, University of Chicago, Chicago IL 60637, USA}
\author{J. Grebel}
\affiliation{Pritzker School of Molecular Engineering, University of Chicago, Chicago IL 60637, USA}
\author{R. G. Povey}
\affiliation{Pritzker School of Molecular Engineering, University of Chicago, Chicago IL 60637, USA}
\affiliation{Department of Physics, University of Chicago, Chicago IL 60637, USA}
\author{K. J. Satzinger}
\altaffiliation[Present address: ]{Google, Santa Barbara CA 93117, USA.}
\affiliation{Department of Physics, University of California, Santa Barbara CA 93106, USA}
\affiliation{Pritzker School of Molecular Engineering, University of Chicago, Chicago IL 60637, USA}
\author{A. N. Cleland}
\affiliation{Pritzker School of Molecular Engineering, University of Chicago, Chicago IL 60637, USA}
\affiliation{Argonne National Laboratory, Argonne IL 60439, USA}

\date{\today}

\begin{abstract}
Phonon modes at microwave frequencies can be cooled to their quantum ground state using conventional cryogenic refrigeration, providing a convenient way to study and manipulate quantum states at the single phonon level. Phonons are of particular interest because mechanical deformations can mediate interactions with a wide range of different quantum systems, including solid-state defects, superconducting qubits, as well as optical photons when using optomechanically-active constructs. Phonons thus hold promise for quantum-focused applications as diverse as sensing, information processing, and communication. Here, we describe a piezoelectric quantum bulk acoustic resonator (QBAR) with a 4.88 GHz resonant frequency that at cryogenic temperatures displays large electromechanical coupling strength combined with a high intrinsic mechanical quality factor $Q_i \approx 4.3 \times 10^4$. Using a recently-developed flip-chip technique, we couple this QBAR resonator to a superconducting qubit on a separate die and demonstrate quantum control of the mechanics in the coupled system. This approach promises a facile and flexible experimental approach to quantum acoustics and hybrid quantum systems.
\end{abstract}

\maketitle

Hybrid quantum systems have attracted significant recent interest, both for applications in quantum information processing \cite{clerk2020,xiang2013,aspelmeyer2014} and in quantum engineering and technology \cite{degen2017,morton2018,kurizki2015}. Quantum acoustics can play an essential role in hybrid quantum systems, as mechanical degrees of freedom can couple to many systems, including superconducting qubits\cite{oconnell2010,gustafsson2014,chu2017,chu2018,bienfait2019,satzinger2018,arrangoiz-arriola2019,sletten2019}, spin ensembles\cite{whiteley2019,chen2019} and optical photons\cite{chan2011,safavi-naeini2012,bochmann2013,vainsencher2016,hong2017,riedinger2018}, and can serve as quantum memories\cite{maccabe2019,ren2019}. On-demand generation of single phonons has been achieved by coupling superconducting qubits via a piezoelectric interaction to film bulk acoustic resonators, to surface acoustic wave resonators and to bulk acoustic resonators \cite{oconnell2010,satzinger2018,chu2018,bienfait2019}. However phonons do not approach the lifetimes of photons in electromagnetic cavities \cite{romanenko2020}. Here we describe one approach that may achieve levels of performance similar to photons while affording simple lithographic fabrication.

\begin{figure}
	\begin{center}
		\includegraphics[width=3.37in]{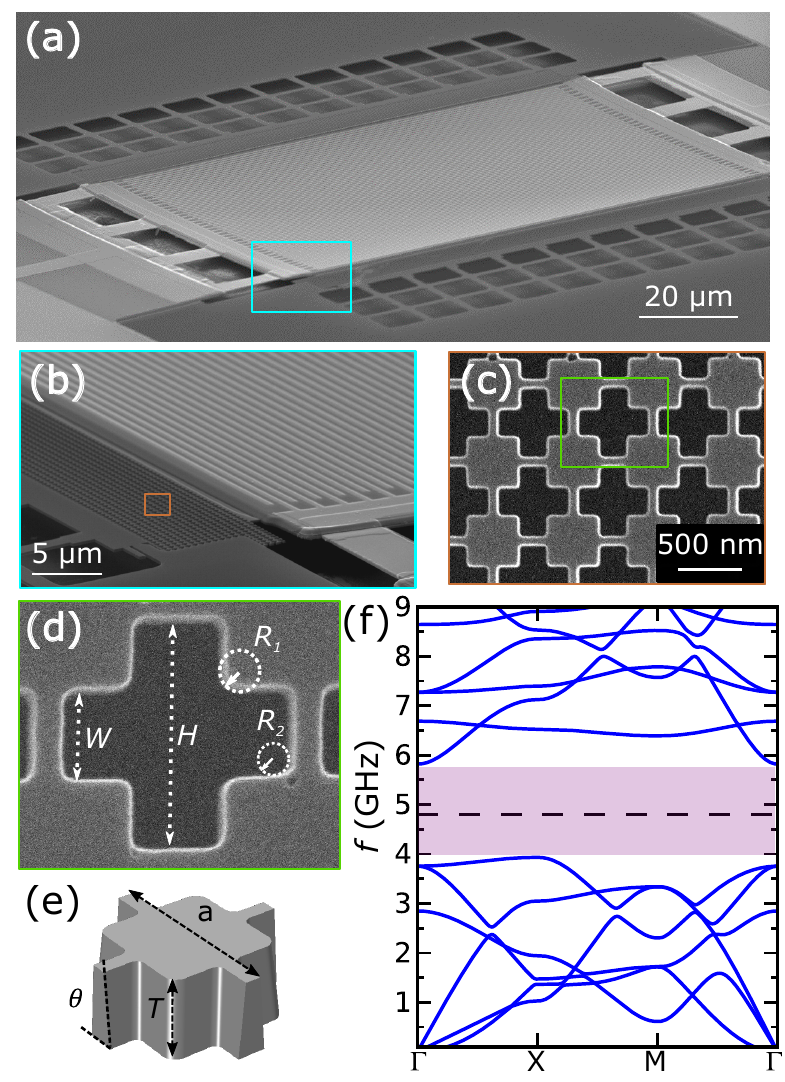}
	\end{center}
	\caption{
		\label{fig:device}
		{Thin film quantum bulk acoustic resonator (QBAR).}
		{(a)} Scanning electron micrograph (SEM) of the resonator structure, an acoustically-resonant membrane $200\times 80$ $\si{{\micro m}^2}$ in area, made of a 220 nm thick single-crystal Si layer supporting a 330 nm thick, $c$-axis oriented piezoelectric AlN layer, supported on either side by an acoustic mirror. Actuation and detection is via an IDT comprising forty Al finger pairs, with alternate fingers connected to one of a pair of wirebond pads (not shown).
		{(b-c)} Details of the structure in (a), showing details of the IDT and the acoustic mirror supports. Each mirror is 15 phononic crystal unit cells wide, on either side of the resonator.
		{(d-e),} Detailed design of the phononic crystal unit cell, with dimension (H, W, R$_1$, R$_2$, $a$, $T$)=($466$, $177$, $40$, $25$, $550$, $220$) nm. The sidewall angle $\theta$ is $\sim$\SI{85}{\degree}.
		{(f)} Finite-element simulation of the band structure for the phononic crystal; dashed line indicates the QBAR resonant frequency; the phononic crystal bandgap is shaded purple.
	}
\end{figure}

Our system comprises a high-$Q$ electromechanical resonator made on one substrate and a superconducting qubit made on a separate substrate, the two coupled using a flip-chip method \cite{satzinger2019}. The mechanical resonator, shown in Fig.~\ref{fig:device}, is a mechanically-suspended bilayer of single-crystal Si with a piezoelectrically-active aluminum nitride (AlN) layer, actuated using an interdigital transducer (IDT) which yields a large electromechanical coupling. The structure is supported by acoustic mirrors\cite{maccabe2019}, giving a high intrinsic mechanical quality factor. The structure exhibits a resonant mechanical mode at 4.88 GHz, making the mechanical quantum ground state accessible by cooling the device to mK temperatures. We term this device a quantum bulk acoustic resonator (QBAR).

The mechanical device is fabricated on a commercial silicon-on-insulator (SOI) wafer with a \SI{220}{\nm} device layer and a \SI{2}{\micro m} buried oxide layer. We first deposit and pattern a \SI{70}{\nm} thick SiO$_x$ stop layer, which protects the active device area's top Si surface from subsequent etching steps. Next, a $c$-axis oriented \SI{330}{\nm} thick aluminum nitride (AlN) piezoelectric layer is deposited by reactive sputter deposition\cite{felmetsger2009}, using conditions that typically yield an in-plane tensile stress below \SI{200}{\MPa}. The AlN film is reactive-ion etched (RIE) using a reflowed photoresist mask, and the exposed underlying SiO$_x$ stop layer removed using buffered HF. To avoid subsequent damage to the AlN, we deposit a \SI{\sim 5}{\nm} SiO$_x$ layer using atomic layer deposition. Phononic crystals are patterned using electron beam lithography, followed by a Cl$_2$/O$_2$ RIE. E-beam lithography defines a PMMA bilayer for lift-off of a \SI{30}{\nm} thick aluminum interdigital transducer (IDT) and ground plane. The wafer then is cut into dies, each having a similar design. The devices are released in HF vapor; an image is shown in Fig~\ref{fig:device}.

\begin{figure}
	\begin{center}
		\includegraphics[width=3.37in]{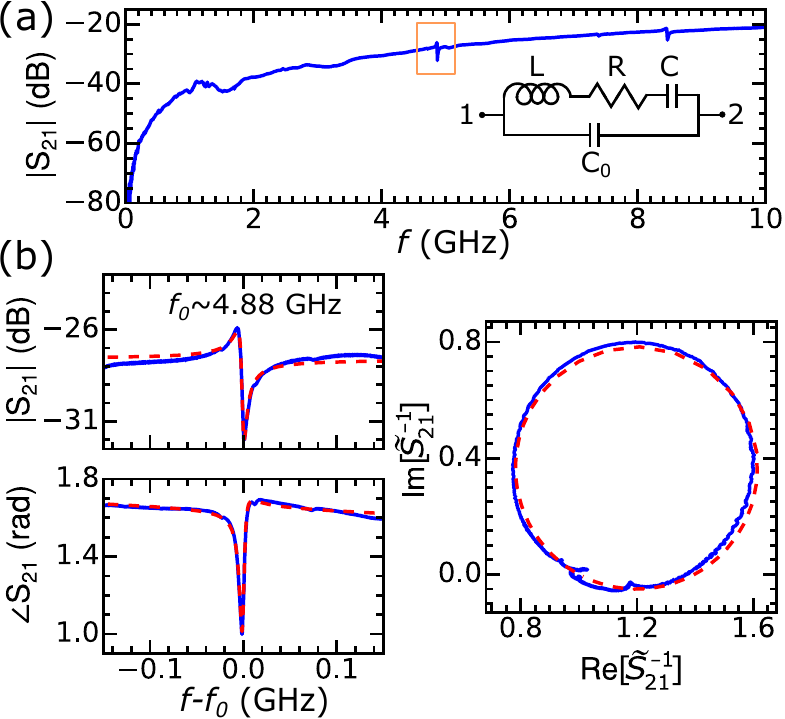}
	\end{center}
	\caption{
		\label{fig:room temp}
		{Room temperature microwave transmission measurements.}
		{(a)} Main panel shows the transmission magnitude $|S_{21}|$ measured by a vector network analyzer (VNA) connected via a microwave probe station. Insert shows equivalent electrical circuit, based on Butterworth-van Dyke model; 1 (2) correspond to VNA ports.
		{(b)} Left: Magnitude $|{S}_{21}|$ and phase $\protect\angle{S}_{21}$ (blue) near the QBAR resonance at $f_0= 4.88$ GHz. Right: Inverse normalized $1/\tilde{S}_{21}$ (blue) in complex plane (horizontal axis: real part, vertical axis: imaginary part). Dashed lines (red) are fits to Eq.~\ref{eqn:s21}.
	}
\end{figure}

The electromechanical resonator is characterized using a calibrated vector network analyzer (VNA), shown in Fig.~\ref{fig:room temp}. We find a strong resonant response at $\omega_0/2\pi = f_0 = 4.88$ GHz, as expected, and fit the response to a Butterworth-van Dyke (BvD) model\cite{oconnell2010,larson2000}. Close to the parallel resonance $f_0 = \sqrt{(C+C_0)/(LCC_0)}/2\pi$, the BvD model has an equivalent impedance $Z_r (f)$ given by
\begin{equation}\label{Zr}
    Z_r(f) \approx \frac{Q_i |Z_1| e^{i\phi}}{1 + 2 i Q_i (f-f_0)/f_0},
\end{equation}
where $Z_1 = \left ( 1 + i 2 \pi f_0 R C - 4 \pi^2 f_0^2 L C \right )/\left (2 \pi f_0 (C+C_0) \right )$, $Q_i = \sqrt{L(C+C_0)/CC_0}/R$ is the internal quality factor and $e^{i\phi}$ is a phase factor. We define the normalized inverse transmission\cite{megrant2012} $\tilde{S}_{21}^{-1}$,
\begin{equation}\label{eqn:s21}
    \tilde{S}_{21}^{-1}(f) \equiv 1 + e^{i\phi}\frac{Q_i}{Q_c}\frac{1}{1 + i 2 Q_i (f-f_0)/f_0},
\end{equation}
where $Q_c = |Z_1|/2Z_0$ is the coupling quality factor and $Z_0 = 50~\Omega$. A fit to the data (Fig.~\ref{fig:room temp}) yields the internal quality factor $Q_i \sim 1.0 \times 10^3$ (measured at room temperature).

\begin{figure}
	\begin{center}
		\includegraphics[width=3.37in]{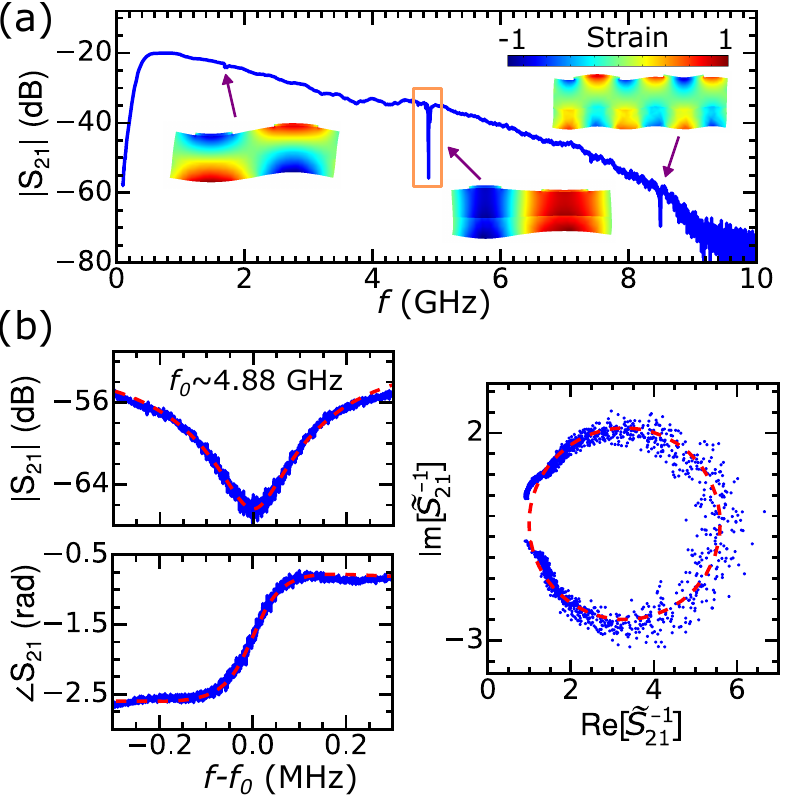}
	\end{center}
	\caption{
		\label{fig:cryo temp}
		{Microwave transmission measurements at $T \sim 80$ mK.}
		{(a)} Transmission magnitude $|S_{21}|$ displays three resonances at 1.69, 4.88, and \SI{8.50}{GHz}, in reasonable agreement with simulations, shown as strain maps (normalized), inset.
		{(b)} Left: Details of the primary resonance at $f_0= 4.88$ GHz, plotted in amplitude and phase versus detuning $f-f_0$. Right: $\tilde{S}_{21}^{-1}$ plotted in the complex plane (blue). Dashed lines (red) are fits to Eq.~\ref{eqn:s21}.
	}
\end{figure}

We characterize the resonator at temperatures below 1 K using an adiabatic demagnetization refrigerator (base temperature $\sim$\SI{60}{mK}). Excitation signals from a VNA pass through a 20 dB attenuator, with the reflection from the device amplified by room-temperature amplifiers with a net gain of 20 dB. Results are displayed in Fig.~\ref{fig:cryo temp}. The resonant frequency remains unchanged from room temperature, while $Q_i$ increases by a factor of 40 to $Q_i \sim 4.3 \times 10^4$. As substrate loss is significantly decreased at cryogenic temperatures, additional resonant modes become detectable, consistent with finite-element simulations, shown in Fig.~\ref{fig:cryo temp}(a).

\begin{figure}
	\begin{center}
		\includegraphics[width=3.37in]{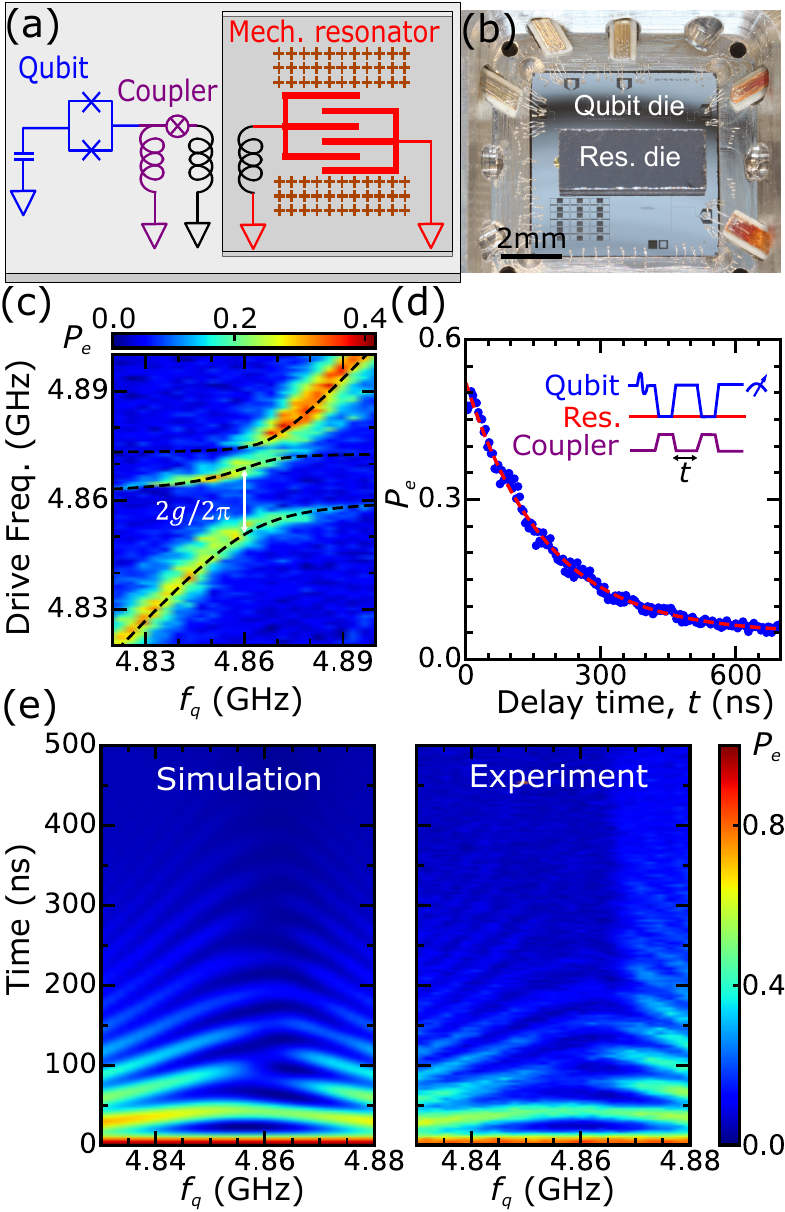}
	\end{center}
	\caption{
		\label{fig:QBAR}
		{Qubit-mediated measurements of mechanical resonator.}
		{(a)} Electrical circuit diagram, showing qubit (blue) and tunable coupler (purple), one arm of which couples inductively (black) to IDT (red). Two acoustic mirrors consist of phononic crystals arrays (brown).
		{(b)} Photograph of flip-chip assembly, comprising a $6\times6$ $\si{{cm}^2}$ qubit die (bottom) and $4\times2$ $\si{{cm}^2}$ resonator die (top).
		{(c)} Qubit spectroscopy, showing excited state probability $P_e$ (color) vs. qubit frequency (horizontal) and microwave pulse frequency (vertical). An avoided-level crossing appears when qubit and resonator are in resonance. Two energy splittings can be observed, the larger corresponding to the primary mechanical mode ($2g/2\pi  \SI{\sim9.6}{\MHz}$), the other a spurious mechanical mode ($2g_{spur}/2\pi  \SI{\sim3.5}{\MHz}$). Dashed lines (black) are fits to a modified Jaynes-Cummings model including two resonant modes.
		{(d)} Phonon lifetime measurement. Inset shows pulse sequence. Main panel shows qubit final excited state probability $P_e$, where the exponential decay is primarily due to the phonon lifetime of 178 ns, as fit by the dashed line (red).
		{(e)} Qubit-resonator Rabi swaps. Probability of the qubit excited state $P_e$ (color scale) is plotted versus qubit frequency $f_q$ (horizontal) and qubit-resonator interaction time (vertical). Coupling strength are $2g/2\pi \SI{\sim 11.2}{\MHz}$ and $2g_{spur}/2\pi  \SI{\sim3.5}{\MHz}$ for primary and spurious mechanical modes, respectively. Left: Simulation results. Right: Experimental results.
	}
\end{figure}

A superconducting qubit is a unique tool to characterize mechanical resonators in the quantum limit \cite{oconnell2010,chu2018,satzinger2018,bienfait2019,bienfait2020}. Here we use a frequency-tunable planar Xmon qubit\cite{koch2007,barends2013} to characterize a QBAR very similar to that measured classically. The qubit is fabricated on a separate sapphire die, with wiring on the two dies including mutual inductive couplers \cite{satzinger2019}; a schematic is shown in Fig.~\ref{fig:QBAR}a. The sapphire and SOI dies are aligned and attached to one another using photoresist, with vertical spacing defined by \SI{\sim 5}{\micro m} thick spacers\cite{satzinger2019}. A flux-tunable coupler element \cite{satzinger2018} is placed between the qubit and its mutual coupling inductance, allowing external flux control of the coupling strength, from zero to a maximum of $2g/2\pi \SI{\sim 11.2}{\MHz}$.

With the coupler off (coupling rate $2g/2 \pi \approx 0$ MHz), we measure the intrinsic qubit $T_1^\mathrm{qb} = \SI{10}{\micro s}$ and $T_{2,Ramsey}^\mathrm{qb} = \SI{1}{\micro s}$, for qubit frequencies ranging from 4.5 to \SI{5.0}{\giga \hertz}, both measured using standard techniques\cite{barends2013}. As we increase the coupling strength from zero, the qubit response includes the resonator and becomes more complex, in particular near the resonator frequency. In Fig.~\ref{fig:QBAR}(c), we show a qubit spectroscopy measurement with the coupler set to a coupling $2g/2\pi = \SI{9.6}{\MHz}$. After setting the qubit frequency (horizontal axis), the qubit is gently excited by a 1$\mu$s excitation microwave tone at the drive frequency (vertical axis), and the qubit excited state probability $P_e$ measured (color scale). The qubit tunes as expected, exhibiting the expected splitting as it crosses the mechanical resonator frequency at $f_r = \SI{4.86}{\giga \hertz}$. There is an additional spurious mode that is weakly coupled to the qubit at \SI{4.87}{\giga \hertz}, with a splitting of about $2g_{spur}/2\pi = \SI{3.5}{\MHz}$. This spurious mode may come from a slight difference between the IDT resonant frequency and that of the QBAR.

We next use the qubit to perform a single-phonon lifetime measurement\cite{oconnell2010}, using the pulse sequence in Fig.~\ref{fig:QBAR}(d). From the decay of $P_e (t)$ with delay $t$, we extract the resonator's energy relaxation time $T_{1,r} = 178\pm2$ ns. This corresponds to a single-phonon quality factor $Q_i \sim (5.43 \pm 0.06) \times 10^3$, slightly smaller than the device measured in Fig.~\ref{fig:cryo temp}.

In Fig.~\ref{fig:QBAR}(e), we display a qubit-resonator Rabi swap, measured as a function of time (vertical axis) and as a function of qubit detuning from the resonator frequency (horizontal axis). A microwave pulse places the qubit in its excited state, and the coupling between qubit and resonator is turned on, initiating the Rabi swap. By measuring the qubit state at different times, we capture the excitation as it is exchanged between qubit and resonator, where as the qubit-resonator detuning increases, the swap rate increases but the amplitude decreases. The spurious mode interferes with this process, generating a non-ideal response, consistent with the spectroscopy measurement. The lifetime of the Rabi swap process is significantly shorter than that measured in Fig.~\ref{fig:QBAR}(d), implying that an unknown additional loss is introduced when we leave the qubit variable coupler on.

We used numerical simulations to support our experimental results. The simulations use a modified Jaynes-Cummings model, where the qubit is modeled as a two-level system coupled to two harmonic oscillators, representing the main and spurious mechanical modes, with different coupling strengths at the frequencies 4.86 and \SI{4.87}{\giga \hertz}, respectively. The avoided-level crossing in Fig.~\ref{fig:QBAR}(c) and the Rabi swap measurement in Fig.~\ref{fig:QBAR}(e) are both supported by this model, from which we extract a $T_{1,r}^\mathrm{spur}$ lifetime for the spurious mode of $~\sim 70$ ns.

In conclusion, we have designed and fabricated a microwave-frequency quantum bulk acoustic resonator with a resonance frequency just below 5 GHz and a single-phonon intrinsic quality factor of $Q_i \sim (5.43\pm0.06) \times 10^3$; a companion device measured with a VNA has $Q_i \sim 4.3 \times 10^4$. These quality factors are roughly 20 times and 200 times higher than previous experiments \cite{oconnell2010}. The piezoelectric construction of the resonator supports a strong electromechanical coupling rate, allowing us to couple it to a superconducting qubit for quantum measurements. This approach holds promise for high quality factor, very small form-factor resonant acoustic cavities operating in the quantum limit, with potential applications to hybrid quantum systems, quantum communication and quantum computing.

\subsection*{Data Availability Statement}
The data that support the findings of this study are available from the corresponding author upon reasonable request.

\subsection*{Acknowledgements}
The authors thank P. J. Duda for fabrication assistance and F. J. Heremans and N. Delegan for help with the AlN sputter system. Devices and experiments were supported by the Air Force Office of Scientific Research and the Army Research Laboratory. K.J.S. was supported by NSF GRFP (NSF DGE-1144085), \'E.D. was supported by LDRD funds from Argonne National Laboratory; A.N.C. was supported in part by the DOE, Office of Basic Energy Sciences. This work was partially supported by the UChicago MRSEC (NSF DMR-1420709) and made use of the Pritzker Nanofabrication Facility, which receives support from SHyNE, a node of the National Science Foundation's National Nanotechnology Coordinated Infrastructure (NSF NNCI-1542205). The authors declare no competing financial interests. Correspondence and requests for materials should be addressed to A. N. Cleland (anc@uchicago.edu).

\bibliographystyle{apsrev4-1}
\bibliography{bibliography_ac}

\begin{thebibliography}{32}%
\makeatletter
\providecommand \@ifxundefined [1]{%
 \@ifx{#1\undefined}
}%
\providecommand \@ifnum [1]{%
 \ifnum #1\expandafter \@firstoftwo
 \else \expandafter \@secondoftwo
 \fi
}%
\providecommand \@ifx [1]{%
 \ifx #1\expandafter \@firstoftwo
 \else \expandafter \@secondoftwo
 \fi
}%
\providecommand \natexlab [1]{#1}%
\providecommand \enquote  [1]{``#1''}%
\providecommand \bibnamefont  [1]{#1}%
\providecommand \bibfnamefont [1]{#1}%
\providecommand \citenamefont [1]{#1}%
\providecommand \href@noop [0]{\@secondoftwo}%
\providecommand \href [0]{\begingroup \@sanitize@url \@href}%
\providecommand \@href[1]{\@@startlink{#1}\@@href}%
\providecommand \@@href[1]{\endgroup#1\@@endlink}%
\providecommand \@sanitize@url [0]{\catcode `\\12\catcode `\$12\catcode
  `\&12\catcode `\#12\catcode `\^12\catcode `\_12\catcode `\%12\relax}%
\providecommand \@@startlink[1]{}%
\providecommand \@@endlink[0]{}%
\providecommand \url  [0]{\begingroup\@sanitize@url \@url }%
\providecommand \@url [1]{\endgroup\@href {#1}{\urlprefix }}%
\providecommand \urlprefix  [0]{URL }%
\providecommand \Eprint [0]{\href }%
\providecommand \doibase [0]{http://dx.doi.org/}%
\providecommand \selectlanguage [0]{\@gobble}%
\providecommand \bibinfo  [0]{\@secondoftwo}%
\providecommand \bibfield  [0]{\@secondoftwo}%
\providecommand \translation [1]{[#1]}%
\providecommand \BibitemOpen [0]{}%
\providecommand \bibitemStop [0]{}%
\providecommand \bibitemNoStop [0]{.\EOS\space}%
\providecommand \EOS [0]{\spacefactor3000\relax}%
\providecommand \BibitemShut  [1]{\csname bibitem#1\endcsname}%
\let\auto@bib@innerbib\@empty
\bibitem [{\citenamefont {Clerk}\ \emph {et~al.}(2020)\citenamefont {Clerk},
  \citenamefont {Lehnert}, \citenamefont {Bertet}, \citenamefont {Petta},\ and\
  \citenamefont {Nakamura}}]{clerk2020}%
  \BibitemOpen
  \bibfield  {author} {\bibinfo {author} {\bibfnamefont {A.~A.}\ \bibnamefont
  {Clerk}}, \bibinfo {author} {\bibfnamefont {K.~W.}\ \bibnamefont {Lehnert}},
  \bibinfo {author} {\bibfnamefont {P.}~\bibnamefont {Bertet}}, \bibinfo
  {author} {\bibfnamefont {J.~R.}\ \bibnamefont {Petta}}, \ and\ \bibinfo
  {author} {\bibfnamefont {Y.}~\bibnamefont {Nakamura}},\ }\href {\doibase
  10.1038/s41567-020-0797-9} {\bibfield  {journal} {\bibinfo  {journal} {Nature
  Physics}\ }\textbf {\bibinfo {volume} {16}},\ \bibinfo {pages} {257}
  (\bibinfo {year} {2020})}\BibitemShut {NoStop}%
\bibitem [{\citenamefont {Xiang}\ \emph {et~al.}(2013)\citenamefont {Xiang},
  \citenamefont {Ashhab}, \citenamefont {You},\ and\ \citenamefont
  {Nori}}]{xiang2013}%
  \BibitemOpen
  \bibfield  {author} {\bibinfo {author} {\bibfnamefont {Z.-L.}\ \bibnamefont
  {Xiang}}, \bibinfo {author} {\bibfnamefont {S.}~\bibnamefont {Ashhab}},
  \bibinfo {author} {\bibfnamefont {J.~Q.}\ \bibnamefont {You}}, \ and\
  \bibinfo {author} {\bibfnamefont {F.}~\bibnamefont {Nori}},\ }\href {\doibase
  10.1103/RevModPhys.85.623} {\bibfield  {journal} {\bibinfo  {journal}
  {Reviews of Modern Physics}\ }\textbf {\bibinfo {volume} {85}},\ \bibinfo
  {pages} {623} (\bibinfo {year} {2013})}\BibitemShut {NoStop}%
\bibitem [{\citenamefont {Aspelmeyer}\ \emph {et~al.}(2014)\citenamefont
  {Aspelmeyer}, \citenamefont {Kippenberg},\ and\ \citenamefont
  {Marquardt}}]{aspelmeyer2014}%
  \BibitemOpen
  \bibfield  {author} {\bibinfo {author} {\bibfnamefont {M.}~\bibnamefont
  {Aspelmeyer}}, \bibinfo {author} {\bibfnamefont {T.~J.}\ \bibnamefont
  {Kippenberg}}, \ and\ \bibinfo {author} {\bibfnamefont {F.}~\bibnamefont
  {Marquardt}},\ }\href {\doibase 10.1103/RevModPhys.86.1391} {\bibfield
  {journal} {\bibinfo  {journal} {Reviews of Modern Physics}\ }\textbf
  {\bibinfo {volume} {86}},\ \bibinfo {pages} {1391} (\bibinfo {year}
  {2014})}\BibitemShut {NoStop}%
\bibitem [{\citenamefont {Degen}\ \emph {et~al.}(2017)\citenamefont {Degen},
  \citenamefont {Reinhard},\ and\ \citenamefont {Cappellaro}}]{degen2017}%
  \BibitemOpen
  \bibfield  {author} {\bibinfo {author} {\bibfnamefont {C.}~\bibnamefont
  {Degen}}, \bibinfo {author} {\bibfnamefont {F.}~\bibnamefont {Reinhard}}, \
  and\ \bibinfo {author} {\bibfnamefont {P.}~\bibnamefont {Cappellaro}},\
  }\href {\doibase 10.1103/RevModPhys.89.035002} {\bibfield  {journal}
  {\bibinfo  {journal} {Reviews of Modern Physics}\ }\textbf {\bibinfo {volume}
  {89}},\ \bibinfo {pages} {035002} (\bibinfo {year} {2017})}\BibitemShut
  {NoStop}%
\bibitem [{\citenamefont {Morton}\ and\ \citenamefont
  {Bertet}(2018)}]{morton2018}%
  \BibitemOpen
  \bibfield  {author} {\bibinfo {author} {\bibfnamefont {J.~J.}\ \bibnamefont
  {Morton}}\ and\ \bibinfo {author} {\bibfnamefont {P.}~\bibnamefont
  {Bertet}},\ }\href {\doibase 10.1016/j.jmr.2017.11.015} {\bibfield  {journal}
  {\bibinfo  {journal} {Journal of Magnetic Resonance}\ }\textbf {\bibinfo
  {volume} {287}},\ \bibinfo {pages} {128} (\bibinfo {year}
  {2018})}\BibitemShut {NoStop}%
\bibitem [{\citenamefont {Kurizki}\ \emph {et~al.}(2015)\citenamefont
  {Kurizki}, \citenamefont {Bertet}, \citenamefont {Kubo}, \citenamefont
  {Mølmer}, \citenamefont {Petrosyan}, \citenamefont {Rabl},\ and\
  \citenamefont {Schmiedmayer}}]{kurizki2015}%
  \BibitemOpen
  \bibfield  {author} {\bibinfo {author} {\bibfnamefont {G.}~\bibnamefont
  {Kurizki}}, \bibinfo {author} {\bibfnamefont {P.}~\bibnamefont {Bertet}},
  \bibinfo {author} {\bibfnamefont {Y.}~\bibnamefont {Kubo}}, \bibinfo {author}
  {\bibfnamefont {K.}~\bibnamefont {Mølmer}}, \bibinfo {author} {\bibfnamefont
  {D.}~\bibnamefont {Petrosyan}}, \bibinfo {author} {\bibfnamefont
  {P.}~\bibnamefont {Rabl}}, \ and\ \bibinfo {author} {\bibfnamefont
  {J.}~\bibnamefont {Schmiedmayer}},\ }\href {\doibase 10.1073/pnas.1419326112}
  {\bibfield  {journal} {\bibinfo  {journal} {Proceedings of the National
  Academy of Sciences}\ }\textbf {\bibinfo {volume} {112}},\ \bibinfo {pages}
  {3866} (\bibinfo {year} {2015})}\BibitemShut {NoStop}%
\bibitem [{\citenamefont {O’Connell}\ \emph {et~al.}(2010)\citenamefont
  {O’Connell}, \citenamefont {Hofheinz}, \citenamefont {Ansmann},
  \citenamefont {Bialczak}, \citenamefont {Lenander}, \citenamefont {Lucero},
  \citenamefont {Neeley}, \citenamefont {Sank}, \citenamefont {Wang},
  \citenamefont {Weides}, \citenamefont {Wenner}, \citenamefont {Martinis},\
  and\ \citenamefont {Cleland}}]{oconnell2010}%
  \BibitemOpen
  \bibfield  {author} {\bibinfo {author} {\bibfnamefont {A.~D.}\ \bibnamefont
  {O’Connell}}, \bibinfo {author} {\bibfnamefont {M.}~\bibnamefont
  {Hofheinz}}, \bibinfo {author} {\bibfnamefont {M.}~\bibnamefont {Ansmann}},
  \bibinfo {author} {\bibfnamefont {R.~C.}\ \bibnamefont {Bialczak}}, \bibinfo
  {author} {\bibfnamefont {M.}~\bibnamefont {Lenander}}, \bibinfo {author}
  {\bibfnamefont {E.}~\bibnamefont {Lucero}}, \bibinfo {author} {\bibfnamefont
  {M.}~\bibnamefont {Neeley}}, \bibinfo {author} {\bibfnamefont
  {D.}~\bibnamefont {Sank}}, \bibinfo {author} {\bibfnamefont {H.}~\bibnamefont
  {Wang}}, \bibinfo {author} {\bibfnamefont {M.}~\bibnamefont {Weides}},
  \bibinfo {author} {\bibfnamefont {J.}~\bibnamefont {Wenner}}, \bibinfo
  {author} {\bibfnamefont {J.~M.}\ \bibnamefont {Martinis}}, \ and\ \bibinfo
  {author} {\bibfnamefont {A.~N.}\ \bibnamefont {Cleland}},\ }\href {\doibase
  10.1038/nature08967} {\bibfield  {journal} {\bibinfo  {journal} {Nature}\
  }\textbf {\bibinfo {volume} {464}},\ \bibinfo {pages} {697} (\bibinfo {year}
  {2010})}\BibitemShut {NoStop}%
\bibitem [{\citenamefont {Gustafsson}\ \emph {et~al.}(2014)\citenamefont
  {Gustafsson}, \citenamefont {Aref}, \citenamefont {Kockum}, \citenamefont
  {Ekstrom}, \citenamefont {Johansson},\ and\ \citenamefont
  {Delsing}}]{gustafsson2014}%
  \BibitemOpen
  \bibfield  {author} {\bibinfo {author} {\bibfnamefont {M.~V.}\ \bibnamefont
  {Gustafsson}}, \bibinfo {author} {\bibfnamefont {T.}~\bibnamefont {Aref}},
  \bibinfo {author} {\bibfnamefont {A.~F.}\ \bibnamefont {Kockum}}, \bibinfo
  {author} {\bibfnamefont {M.~K.}\ \bibnamefont {Ekstrom}}, \bibinfo {author}
  {\bibfnamefont {G.}~\bibnamefont {Johansson}}, \ and\ \bibinfo {author}
  {\bibfnamefont {P.}~\bibnamefont {Delsing}},\ }\href {\doibase
  10.1126/science.1257219} {\bibfield  {journal} {\bibinfo  {journal}
  {Science}\ }\textbf {\bibinfo {volume} {346}},\ \bibinfo {pages} {207}
  (\bibinfo {year} {2014})}\BibitemShut {NoStop}%
\bibitem [{\citenamefont {Chu}\ \emph {et~al.}(2017)\citenamefont {Chu},
  \citenamefont {Kharel}, \citenamefont {Renninger}, \citenamefont {Burkhart},
  \citenamefont {Frunzio}, \citenamefont {Rakich},\ and\ \citenamefont
  {Schoelkopf}}]{chu2017}%
  \BibitemOpen
  \bibfield  {author} {\bibinfo {author} {\bibfnamefont {Y.}~\bibnamefont
  {Chu}}, \bibinfo {author} {\bibfnamefont {P.}~\bibnamefont {Kharel}},
  \bibinfo {author} {\bibfnamefont {W.~H.}\ \bibnamefont {Renninger}}, \bibinfo
  {author} {\bibfnamefont {L.~D.}\ \bibnamefont {Burkhart}}, \bibinfo {author}
  {\bibfnamefont {L.}~\bibnamefont {Frunzio}}, \bibinfo {author} {\bibfnamefont
  {P.~T.}\ \bibnamefont {Rakich}}, \ and\ \bibinfo {author} {\bibfnamefont
  {R.~J.}\ \bibnamefont {Schoelkopf}},\ }\href {\doibase
  10.1126/science.aao1511} {\bibfield  {journal} {\bibinfo  {journal}
  {Science}\ }\textbf {\bibinfo {volume} {358}},\ \bibinfo {pages} {199}
  (\bibinfo {year} {2017})}\BibitemShut {NoStop}%
\bibitem [{\citenamefont {Chu}\ \emph {et~al.}(2018)\citenamefont {Chu},
  \citenamefont {Kharel}, \citenamefont {Yoon}, \citenamefont {Frunzio},
  \citenamefont {Rakich},\ and\ \citenamefont {Schoelkopf}}]{chu2018}%
  \BibitemOpen
  \bibfield  {author} {\bibinfo {author} {\bibfnamefont {Y.}~\bibnamefont
  {Chu}}, \bibinfo {author} {\bibfnamefont {P.}~\bibnamefont {Kharel}},
  \bibinfo {author} {\bibfnamefont {T.}~\bibnamefont {Yoon}}, \bibinfo {author}
  {\bibfnamefont {L.}~\bibnamefont {Frunzio}}, \bibinfo {author} {\bibfnamefont
  {P.~T.}\ \bibnamefont {Rakich}}, \ and\ \bibinfo {author} {\bibfnamefont
  {R.~J.}\ \bibnamefont {Schoelkopf}},\ }\href {\doibase
  10.1038/s41586-018-0717-7} {\bibfield  {journal} {\bibinfo  {journal}
  {Nature}\ }\textbf {\bibinfo {volume} {563}},\ \bibinfo {pages} {666}
  (\bibinfo {year} {2018})},\ \bibinfo {note} {arXiv: 1804.07426}\BibitemShut
  {NoStop}%
\bibitem [{\citenamefont {Bienfait}\ \emph {et~al.}(2019)\citenamefont
  {Bienfait}, \citenamefont {Satzinger}, \citenamefont {Zhong}, \citenamefont
  {Chang}, \citenamefont {Chou}, \citenamefont {Conner}, \citenamefont {Dumur},
  \citenamefont {Grebel}, \citenamefont {Peairs}, \citenamefont {Povey},\ and\
  \citenamefont {Cleland}}]{bienfait2019}%
  \BibitemOpen
  \bibfield  {author} {\bibinfo {author} {\bibfnamefont {A.}~\bibnamefont
  {Bienfait}}, \bibinfo {author} {\bibfnamefont {K.~J.}\ \bibnamefont
  {Satzinger}}, \bibinfo {author} {\bibfnamefont {Y.~P.}\ \bibnamefont
  {Zhong}}, \bibinfo {author} {\bibfnamefont {H.-S.}\ \bibnamefont {Chang}},
  \bibinfo {author} {\bibfnamefont {M.-H.}\ \bibnamefont {Chou}}, \bibinfo
  {author} {\bibfnamefont {C.~R.}\ \bibnamefont {Conner}}, \bibinfo {author}
  {\bibfnamefont {.}~\bibnamefont {Dumur}}, \bibinfo {author} {\bibfnamefont
  {J.}~\bibnamefont {Grebel}}, \bibinfo {author} {\bibfnamefont {G.~A.}\
  \bibnamefont {Peairs}}, \bibinfo {author} {\bibfnamefont {R.~G.}\
  \bibnamefont {Povey}}, \ and\ \bibinfo {author} {\bibfnamefont {A.~N.}\
  \bibnamefont {Cleland}},\ }\href {\doibase 10.1126/science.aaw8415}
  {\bibfield  {journal} {\bibinfo  {journal} {Science}\ }\textbf {\bibinfo
  {volume} {364}},\ \bibinfo {pages} {368} (\bibinfo {year}
  {2019})}\BibitemShut {NoStop}%
\bibitem [{\citenamefont {Satzinger}\ \emph {et~al.}(2018)\citenamefont
  {Satzinger}, \citenamefont {Zhong}, \citenamefont {Chang}, \citenamefont
  {Peairs}, \citenamefont {Bienfait}, \citenamefont {Chou}, \citenamefont
  {Cleland}, \citenamefont {Conner}, \citenamefont {Dumur}, \citenamefont
  {Grebel}, \citenamefont {Gutierrez}, \citenamefont {November}, \citenamefont
  {Povey}, \citenamefont {Whiteley}, \citenamefont {Awschalom}, \citenamefont
  {Schuster},\ and\ \citenamefont {Cleland}}]{satzinger2018}%
  \BibitemOpen
  \bibfield  {author} {\bibinfo {author} {\bibfnamefont {K.~J.}\ \bibnamefont
  {Satzinger}}, \bibinfo {author} {\bibfnamefont {Y.~P.}\ \bibnamefont
  {Zhong}}, \bibinfo {author} {\bibfnamefont {H.-S.}\ \bibnamefont {Chang}},
  \bibinfo {author} {\bibfnamefont {G.~A.}\ \bibnamefont {Peairs}}, \bibinfo
  {author} {\bibfnamefont {A.}~\bibnamefont {Bienfait}}, \bibinfo {author}
  {\bibfnamefont {M.-H.}\ \bibnamefont {Chou}}, \bibinfo {author}
  {\bibfnamefont {A.~Y.}\ \bibnamefont {Cleland}}, \bibinfo {author}
  {\bibfnamefont {C.~R.}\ \bibnamefont {Conner}}, \bibinfo {author}
  {\bibfnamefont {.}~\bibnamefont {Dumur}}, \bibinfo {author} {\bibfnamefont
  {J.}~\bibnamefont {Grebel}}, \bibinfo {author} {\bibfnamefont
  {I.}~\bibnamefont {Gutierrez}}, \bibinfo {author} {\bibfnamefont {B.~H.}\
  \bibnamefont {November}}, \bibinfo {author} {\bibfnamefont {R.~G.}\
  \bibnamefont {Povey}}, \bibinfo {author} {\bibfnamefont {S.~J.}\ \bibnamefont
  {Whiteley}}, \bibinfo {author} {\bibfnamefont {D.~D.}\ \bibnamefont
  {Awschalom}}, \bibinfo {author} {\bibfnamefont {D.~I.}\ \bibnamefont
  {Schuster}}, \ and\ \bibinfo {author} {\bibfnamefont {A.~N.}\ \bibnamefont
  {Cleland}},\ }\href {\doibase 10.1038/s41586-018-0719-5} {\bibfield
  {journal} {\bibinfo  {journal} {Nature}\ }\textbf {\bibinfo {volume} {563}},\
  \bibinfo {pages} {661} (\bibinfo {year} {2018})}\BibitemShut {NoStop}%
\bibitem [{\citenamefont {Arrangoiz-Arriola}\ \emph {et~al.}(2019)\citenamefont
  {Arrangoiz-Arriola}, \citenamefont {Wollack}, \citenamefont {Wang},
  \citenamefont {Pechal}, \citenamefont {Jiang}, \citenamefont {McKenna},
  \citenamefont {Witmer},\ and\ \citenamefont
  {Safavi-Naeini}}]{arrangoiz-arriola2019}%
  \BibitemOpen
  \bibfield  {author} {\bibinfo {author} {\bibfnamefont {P.}~\bibnamefont
  {Arrangoiz-Arriola}}, \bibinfo {author} {\bibfnamefont {E.~A.}\ \bibnamefont
  {Wollack}}, \bibinfo {author} {\bibfnamefont {Z.}~\bibnamefont {Wang}},
  \bibinfo {author} {\bibfnamefont {M.}~\bibnamefont {Pechal}}, \bibinfo
  {author} {\bibfnamefont {W.}~\bibnamefont {Jiang}}, \bibinfo {author}
  {\bibfnamefont {T.~P.}\ \bibnamefont {McKenna}}, \bibinfo {author}
  {\bibfnamefont {J.~D.}\ \bibnamefont {Witmer}}, \ and\ \bibinfo {author}
  {\bibfnamefont {A.~H.}\ \bibnamefont {Safavi-Naeini}},\ }\href {\doibase
  10.1038/s41586-019-1386-x} {\bibfield  {journal} {\bibinfo  {journal}
  {Nature}\ }\textbf {\bibinfo {volume} {571}},\ \bibinfo {pages} {537}
  (\bibinfo {year} {2019})},\ \bibinfo {note} {arXiv: 1902.04681}\BibitemShut
  {NoStop}%
\bibitem [{\citenamefont {Sletten}\ \emph {et~al.}(2019)\citenamefont
  {Sletten}, \citenamefont {Moores}, \citenamefont {Viennot},\ and\
  \citenamefont {Lehnert}}]{sletten2019}%
  \BibitemOpen
  \bibfield  {author} {\bibinfo {author} {\bibfnamefont {L.}~\bibnamefont
  {Sletten}}, \bibinfo {author} {\bibfnamefont {B.}~\bibnamefont {Moores}},
  \bibinfo {author} {\bibfnamefont {J.}~\bibnamefont {Viennot}}, \ and\
  \bibinfo {author} {\bibfnamefont {K.}~\bibnamefont {Lehnert}},\ }\href
  {\doibase 10.1103/PhysRevX.9.021056} {\bibfield  {journal} {\bibinfo
  {journal} {Physical Review X}\ }\textbf {\bibinfo {volume} {9}},\ \bibinfo
  {pages} {021056} (\bibinfo {year} {2019})}\BibitemShut {NoStop}%
\bibitem [{\citenamefont {Whiteley}\ \emph {et~al.}(2019)\citenamefont
  {Whiteley}, \citenamefont {Wolfowicz}, \citenamefont {Anderson},
  \citenamefont {Bourassa}, \citenamefont {Ma}, \citenamefont {Ye},
  \citenamefont {Koolstra}, \citenamefont {Satzinger}, \citenamefont {Holt},
  \citenamefont {Heremans}, \citenamefont {Cleland}, \citenamefont {Schuster},
  \citenamefont {Galli},\ and\ \citenamefont {Awschalom}}]{whiteley2019}%
  \BibitemOpen
  \bibfield  {author} {\bibinfo {author} {\bibfnamefont {S.~J.}\ \bibnamefont
  {Whiteley}}, \bibinfo {author} {\bibfnamefont {G.}~\bibnamefont {Wolfowicz}},
  \bibinfo {author} {\bibfnamefont {C.~P.}\ \bibnamefont {Anderson}}, \bibinfo
  {author} {\bibfnamefont {A.}~\bibnamefont {Bourassa}}, \bibinfo {author}
  {\bibfnamefont {H.}~\bibnamefont {Ma}}, \bibinfo {author} {\bibfnamefont
  {M.}~\bibnamefont {Ye}}, \bibinfo {author} {\bibfnamefont {G.}~\bibnamefont
  {Koolstra}}, \bibinfo {author} {\bibfnamefont {K.~J.}\ \bibnamefont
  {Satzinger}}, \bibinfo {author} {\bibfnamefont {M.~V.}\ \bibnamefont {Holt}},
  \bibinfo {author} {\bibfnamefont {F.~J.}\ \bibnamefont {Heremans}}, \bibinfo
  {author} {\bibfnamefont {A.~N.}\ \bibnamefont {Cleland}}, \bibinfo {author}
  {\bibfnamefont {D.~I.}\ \bibnamefont {Schuster}}, \bibinfo {author}
  {\bibfnamefont {G.}~\bibnamefont {Galli}}, \ and\ \bibinfo {author}
  {\bibfnamefont {D.~D.}\ \bibnamefont {Awschalom}},\ }\href {\doibase
  10.1038/s41567-019-0420-0} {\bibfield  {journal} {\bibinfo  {journal} {Nature
  Physics}\ }\textbf {\bibinfo {volume} {15}},\ \bibinfo {pages} {490}
  (\bibinfo {year} {2019})}\BibitemShut {NoStop}%
\bibitem [{\citenamefont {Chen}\ \emph {et~al.}(2019)\citenamefont {Chen},
  \citenamefont {Opondo}, \citenamefont {Jiang}, \citenamefont {MacQuarrie},
  \citenamefont {Daveau}, \citenamefont {Bhave},\ and\ \citenamefont
  {Fuchs}}]{chen2019}%
  \BibitemOpen
  \bibfield  {author} {\bibinfo {author} {\bibfnamefont {H.}~\bibnamefont
  {Chen}}, \bibinfo {author} {\bibfnamefont {N.~F.}\ \bibnamefont {Opondo}},
  \bibinfo {author} {\bibfnamefont {B.}~\bibnamefont {Jiang}}, \bibinfo
  {author} {\bibfnamefont {E.~R.}\ \bibnamefont {MacQuarrie}}, \bibinfo
  {author} {\bibfnamefont {R.~S.}\ \bibnamefont {Daveau}}, \bibinfo {author}
  {\bibfnamefont {S.~A.}\ \bibnamefont {Bhave}}, \ and\ \bibinfo {author}
  {\bibfnamefont {G.~D.}\ \bibnamefont {Fuchs}},\ }\href {\doibase
  10.1021/acs.nanolett.9b02430} {\bibfield  {journal} {\bibinfo  {journal}
  {Nano Letters}\ }\textbf {\bibinfo {volume} {19}},\ \bibinfo {pages} {7021}
  (\bibinfo {year} {2019})}\BibitemShut {NoStop}%
\bibitem [{\citenamefont {Chan}\ \emph {et~al.}(2011)\citenamefont {Chan},
  \citenamefont {Alegre}, \citenamefont {Safavi-Naeini}, \citenamefont {Hill},
  \citenamefont {Krause}, \citenamefont {Gröblacher}, \citenamefont
  {Aspelmeyer},\ and\ \citenamefont {Painter}}]{chan2011}%
  \BibitemOpen
  \bibfield  {author} {\bibinfo {author} {\bibfnamefont {J.}~\bibnamefont
  {Chan}}, \bibinfo {author} {\bibfnamefont {T.~P.~M.}\ \bibnamefont {Alegre}},
  \bibinfo {author} {\bibfnamefont {A.~H.}\ \bibnamefont {Safavi-Naeini}},
  \bibinfo {author} {\bibfnamefont {J.~T.}\ \bibnamefont {Hill}}, \bibinfo
  {author} {\bibfnamefont {A.}~\bibnamefont {Krause}}, \bibinfo {author}
  {\bibfnamefont {S.}~\bibnamefont {Gröblacher}}, \bibinfo {author}
  {\bibfnamefont {M.}~\bibnamefont {Aspelmeyer}}, \ and\ \bibinfo {author}
  {\bibfnamefont {O.}~\bibnamefont {Painter}},\ }\href {\doibase
  10.1038/nature10461} {\bibfield  {journal} {\bibinfo  {journal} {Nature}\
  }\textbf {\bibinfo {volume} {478}},\ \bibinfo {pages} {89} (\bibinfo {year}
  {2011})}\BibitemShut {NoStop}%
\bibitem [{\citenamefont {Safavi-Naeini}\ \emph {et~al.}(2012)\citenamefont
  {Safavi-Naeini}, \citenamefont {Chan}, \citenamefont {Hill}, \citenamefont
  {Alegre}, \citenamefont {Krause},\ and\ \citenamefont
  {Painter}}]{safavi-naeini2012}%
  \BibitemOpen
  \bibfield  {author} {\bibinfo {author} {\bibfnamefont {A.~H.}\ \bibnamefont
  {Safavi-Naeini}}, \bibinfo {author} {\bibfnamefont {J.}~\bibnamefont {Chan}},
  \bibinfo {author} {\bibfnamefont {J.~T.}\ \bibnamefont {Hill}}, \bibinfo
  {author} {\bibfnamefont {T.~P.~M.}\ \bibnamefont {Alegre}}, \bibinfo {author}
  {\bibfnamefont {A.}~\bibnamefont {Krause}}, \ and\ \bibinfo {author}
  {\bibfnamefont {O.}~\bibnamefont {Painter}},\ }\href {\doibase
  10.1103/PhysRevLett.108.033602} {\bibfield  {journal} {\bibinfo  {journal}
  {Physical Review Letters}\ }\textbf {\bibinfo {volume} {108}},\ \bibinfo
  {pages} {033602} (\bibinfo {year} {2012})}\BibitemShut {NoStop}%
\bibitem [{\citenamefont {Bochmann}\ \emph {et~al.}(2013)\citenamefont
  {Bochmann}, \citenamefont {Vainsencher}, \citenamefont {Awschalom},\ and\
  \citenamefont {Cleland}}]{bochmann2013}%
  \BibitemOpen
  \bibfield  {author} {\bibinfo {author} {\bibfnamefont {J.}~\bibnamefont
  {Bochmann}}, \bibinfo {author} {\bibfnamefont {A.}~\bibnamefont
  {Vainsencher}}, \bibinfo {author} {\bibfnamefont {D.~D.}\ \bibnamefont
  {Awschalom}}, \ and\ \bibinfo {author} {\bibfnamefont {A.~N.}\ \bibnamefont
  {Cleland}},\ }\href@noop {} {\bibfield  {journal} {\bibinfo  {journal}
  {Nature Physics}\ }\textbf {\bibinfo {volume} {9}},\ \bibinfo {pages} {712}
  (\bibinfo {year} {2013})}\BibitemShut {NoStop}%
\bibitem [{\citenamefont {Vainsencher}\ \emph {et~al.}(2016)\citenamefont
  {Vainsencher}, \citenamefont {Satzinger}, \citenamefont {Peairs},\ and\
  \citenamefont {Cleland}}]{vainsencher2016}%
  \BibitemOpen
  \bibfield  {author} {\bibinfo {author} {\bibfnamefont {A.}~\bibnamefont
  {Vainsencher}}, \bibinfo {author} {\bibfnamefont {K.}~\bibnamefont
  {Satzinger}}, \bibinfo {author} {\bibfnamefont {G.}~\bibnamefont {Peairs}}, \
  and\ \bibinfo {author} {\bibfnamefont {A.}~\bibnamefont {Cleland}},\
  }\href@noop {} {\bibfield  {journal} {\bibinfo  {journal} {Applied Physics
  Letters}\ }\textbf {\bibinfo {volume} {109}},\ \bibinfo {pages} {033107}
  (\bibinfo {year} {2016})}\BibitemShut {NoStop}%
\bibitem [{\citenamefont {Hong}\ \emph {et~al.}(2017)\citenamefont {Hong},
  \citenamefont {Riedinger}, \citenamefont {Marinkovic}, \citenamefont
  {Wallucks}, \citenamefont {Hofer}, \citenamefont {Norte}, \citenamefont
  {Aspelmeyer},\ and\ \citenamefont {Gröblacher}}]{hong2017}%
  \BibitemOpen
  \bibfield  {author} {\bibinfo {author} {\bibfnamefont {S.}~\bibnamefont
  {Hong}}, \bibinfo {author} {\bibfnamefont {R.}~\bibnamefont {Riedinger}},
  \bibinfo {author} {\bibfnamefont {I.}~\bibnamefont {Marinkovic}}, \bibinfo
  {author} {\bibfnamefont {A.}~\bibnamefont {Wallucks}}, \bibinfo {author}
  {\bibfnamefont {S.~G.}\ \bibnamefont {Hofer}}, \bibinfo {author}
  {\bibfnamefont {R.~A.}\ \bibnamefont {Norte}}, \bibinfo {author}
  {\bibfnamefont {M.}~\bibnamefont {Aspelmeyer}}, \ and\ \bibinfo {author}
  {\bibfnamefont {S.}~\bibnamefont {Gröblacher}},\ }\href {\doibase
  10.1126/science.aan7939} {\bibfield  {journal} {\bibinfo  {journal}
  {Science}\ }\textbf {\bibinfo {volume} {358}},\ \bibinfo {pages} {203}
  (\bibinfo {year} {2017})},\ \bibinfo {note} {arXiv: 1706.03777}\BibitemShut
  {NoStop}%
\bibitem [{\citenamefont {Riedinger}\ \emph {et~al.}(2018)\citenamefont
  {Riedinger}, \citenamefont {Wallucks}, \citenamefont {Marinkovic},
  \citenamefont {Löschnauer}, \citenamefont {Aspelmeyer}, \citenamefont
  {Hong},\ and\ \citenamefont {Gröblacher}}]{riedinger2018}%
  \BibitemOpen
  \bibfield  {author} {\bibinfo {author} {\bibfnamefont {R.}~\bibnamefont
  {Riedinger}}, \bibinfo {author} {\bibfnamefont {A.}~\bibnamefont {Wallucks}},
  \bibinfo {author} {\bibfnamefont {I.}~\bibnamefont {Marinkovic}}, \bibinfo
  {author} {\bibfnamefont {C.}~\bibnamefont {Löschnauer}}, \bibinfo {author}
  {\bibfnamefont {M.}~\bibnamefont {Aspelmeyer}}, \bibinfo {author}
  {\bibfnamefont {S.}~\bibnamefont {Hong}}, \ and\ \bibinfo {author}
  {\bibfnamefont {S.}~\bibnamefont {Gröblacher}},\ }\href {\doibase
  10.1038/s41586-018-0036-z} {\bibfield  {journal} {\bibinfo  {journal}
  {Nature}\ }\textbf {\bibinfo {volume} {556}},\ \bibinfo {pages} {473}
  (\bibinfo {year} {2018})},\ \bibinfo {note} {tex.ids: riedinger2018a arXiv:
  1710.11147}\BibitemShut {NoStop}%
\bibitem [{\citenamefont {MacCabe}\ \emph {et~al.}(2019)\citenamefont
  {MacCabe}, \citenamefont {Ren}, \citenamefont {Luo}, \citenamefont {Cohen},
  \citenamefont {Zhou}, \citenamefont {Sipahigil}, \citenamefont
  {Mirhosseini},\ and\ \citenamefont {Painter}}]{maccabe2019}%
  \BibitemOpen
  \bibfield  {author} {\bibinfo {author} {\bibfnamefont {G.~S.}\ \bibnamefont
  {MacCabe}}, \bibinfo {author} {\bibfnamefont {H.}~\bibnamefont {Ren}},
  \bibinfo {author} {\bibfnamefont {J.}~\bibnamefont {Luo}}, \bibinfo {author}
  {\bibfnamefont {J.~D.}\ \bibnamefont {Cohen}}, \bibinfo {author}
  {\bibfnamefont {H.}~\bibnamefont {Zhou}}, \bibinfo {author} {\bibfnamefont
  {A.}~\bibnamefont {Sipahigil}}, \bibinfo {author} {\bibfnamefont
  {M.}~\bibnamefont {Mirhosseini}}, \ and\ \bibinfo {author} {\bibfnamefont
  {O.}~\bibnamefont {Painter}},\ }\href {http://arxiv.org/abs/1901.04129}
  {\bibfield  {journal} {\bibinfo  {journal} {arXiv:1901.04129 [cond-mat,
  physics:quant-ph]}\ } (\bibinfo {year} {2019})},\ \bibinfo {note} {arXiv:
  1901.04129}\BibitemShut {NoStop}%
\bibitem [{\citenamefont {Ren}\ \emph {et~al.}(2019)\citenamefont {Ren},
  \citenamefont {Matheny}, \citenamefont {MacCabe}, \citenamefont {Luo},
  \citenamefont {Pfeifer}, \citenamefont {Mirhosseini},\ and\ \citenamefont
  {Painter}}]{ren2019}%
  \BibitemOpen
  \bibfield  {author} {\bibinfo {author} {\bibfnamefont {H.}~\bibnamefont
  {Ren}}, \bibinfo {author} {\bibfnamefont {M.~H.}\ \bibnamefont {Matheny}},
  \bibinfo {author} {\bibfnamefont {G.~S.}\ \bibnamefont {MacCabe}}, \bibinfo
  {author} {\bibfnamefont {J.}~\bibnamefont {Luo}}, \bibinfo {author}
  {\bibfnamefont {H.}~\bibnamefont {Pfeifer}}, \bibinfo {author} {\bibfnamefont
  {M.}~\bibnamefont {Mirhosseini}}, \ and\ \bibinfo {author} {\bibfnamefont
  {O.}~\bibnamefont {Painter}},\ }\href {http://arxiv.org/abs/1910.02873}
  {\bibfield  {journal} {\bibinfo  {journal} {arXiv:1910.02873 [physics,
  physics:quant-ph]}\ } (\bibinfo {year} {2019})},\ \bibinfo {note} {arXiv:
  1910.02873}\BibitemShut {NoStop}%
\bibitem [{\citenamefont {Romanenko}\ \emph {et~al.}(2020)\citenamefont
  {Romanenko}, \citenamefont {Pilipenko}, \citenamefont {Zorzetti},
  \citenamefont {Frolov}, \citenamefont {Awida}, \citenamefont {Belomestnykh},
  \citenamefont {Posen},\ and\ \citenamefont {Grassellino}}]{romanenko2020}%
  \BibitemOpen
  \bibfield  {author} {\bibinfo {author} {\bibfnamefont {A.}~\bibnamefont
  {Romanenko}}, \bibinfo {author} {\bibfnamefont {R.}~\bibnamefont
  {Pilipenko}}, \bibinfo {author} {\bibfnamefont {S.}~\bibnamefont {Zorzetti}},
  \bibinfo {author} {\bibfnamefont {D.}~\bibnamefont {Frolov}}, \bibinfo
  {author} {\bibfnamefont {M.}~\bibnamefont {Awida}}, \bibinfo {author}
  {\bibfnamefont {S.}~\bibnamefont {Belomestnykh}}, \bibinfo {author}
  {\bibfnamefont {S.}~\bibnamefont {Posen}}, \ and\ \bibinfo {author}
  {\bibfnamefont {A.}~\bibnamefont {Grassellino}},\ }\href@noop {} {\bibfield
  {journal} {\bibinfo  {journal} {Physical Review Applied}\ }\textbf {\bibinfo
  {volume} {13}},\ \bibinfo {pages} {034032} (\bibinfo {year}
  {2020})}\BibitemShut {NoStop}%
\bibitem [{\citenamefont {Satzinger}\ \emph {et~al.}(2019)\citenamefont
  {Satzinger}, \citenamefont {Conner}, \citenamefont {Bienfait}, \citenamefont
  {Chang}, \citenamefont {Chou}, \citenamefont {Cleland}, \citenamefont
  {Dumur}, \citenamefont {Grebel}, \citenamefont {Peairs}, \citenamefont
  {Povey}, \citenamefont {Whiteley}, \citenamefont {Zhong}, \citenamefont
  {Awschalom}, \citenamefont {Schuster},\ and\ \citenamefont
  {Cleland}}]{satzinger2019}%
  \BibitemOpen
  \bibfield  {author} {\bibinfo {author} {\bibfnamefont {K.~J.}\ \bibnamefont
  {Satzinger}}, \bibinfo {author} {\bibfnamefont {C.~R.}\ \bibnamefont
  {Conner}}, \bibinfo {author} {\bibfnamefont {A.}~\bibnamefont {Bienfait}},
  \bibinfo {author} {\bibfnamefont {H.-S.}\ \bibnamefont {Chang}}, \bibinfo
  {author} {\bibfnamefont {M.-H.}\ \bibnamefont {Chou}}, \bibinfo {author}
  {\bibfnamefont {A.~Y.}\ \bibnamefont {Cleland}}, \bibinfo {author}
  {\bibfnamefont {.}~\bibnamefont {Dumur}}, \bibinfo {author} {\bibfnamefont
  {J.}~\bibnamefont {Grebel}}, \bibinfo {author} {\bibfnamefont {G.~A.}\
  \bibnamefont {Peairs}}, \bibinfo {author} {\bibfnamefont {R.~G.}\
  \bibnamefont {Povey}}, \bibinfo {author} {\bibfnamefont {S.~J.}\ \bibnamefont
  {Whiteley}}, \bibinfo {author} {\bibfnamefont {Y.~P.}\ \bibnamefont {Zhong}},
  \bibinfo {author} {\bibfnamefont {D.~D.}\ \bibnamefont {Awschalom}}, \bibinfo
  {author} {\bibfnamefont {D.~I.}\ \bibnamefont {Schuster}}, \ and\ \bibinfo
  {author} {\bibfnamefont {A.~N.}\ \bibnamefont {Cleland}},\ }\href {\doibase
  10.1063/1.5089888} {\bibfield  {journal} {\bibinfo  {journal} {Applied
  Physics Letters}\ }\textbf {\bibinfo {volume} {114}},\ \bibinfo {pages}
  {173501} (\bibinfo {year} {2019})}\BibitemShut {NoStop}%
\bibitem [{\citenamefont {Felmetsger}\ \emph {et~al.}(2009)\citenamefont
  {Felmetsger}, \citenamefont {Laptev},\ and\ \citenamefont
  {Tanner}}]{felmetsger2009}%
  \BibitemOpen
  \bibfield  {author} {\bibinfo {author} {\bibfnamefont {V.}~\bibnamefont
  {Felmetsger}}, \bibinfo {author} {\bibfnamefont {P.}~\bibnamefont {Laptev}},
  \ and\ \bibinfo {author} {\bibfnamefont {S.}~\bibnamefont {Tanner}},\
  }\href@noop {} {\bibfield  {journal} {\bibinfo  {journal} {Journal of Vacuum
  Science \& Technology A: Vacuum, Surfaces, and Films}\ }\textbf {\bibinfo
  {volume} {27}},\ \bibinfo {pages} {417} (\bibinfo {year} {2009})}\BibitemShut
  {NoStop}%
\bibitem [{\citenamefont {Larson}\ \emph {et~al.}(2000)\citenamefont {Larson},
  \citenamefont {Ruby}, \citenamefont {Bradley}, \citenamefont {Wen},
  \citenamefont {Kok},\ and\ \citenamefont {Chien}}]{larson2000}%
  \BibitemOpen
  \bibfield  {author} {\bibinfo {author} {\bibfnamefont {J.~D.}\ \bibnamefont
  {Larson}}, \bibinfo {author} {\bibfnamefont {J.}~\bibnamefont {Ruby}},
  \bibinfo {author} {\bibfnamefont {R.}~\bibnamefont {Bradley}}, \bibinfo
  {author} {\bibfnamefont {J.}~\bibnamefont {Wen}}, \bibinfo {author}
  {\bibfnamefont {S.-L.}\ \bibnamefont {Kok}}, \ and\ \bibinfo {author}
  {\bibfnamefont {A.}~\bibnamefont {Chien}},\ }\bibfield  {booktitle} {\emph
  {\bibinfo {booktitle} {2000 IEEE Ultrasonics Symposium. Proceedings. An
  International Symposium (Cat. No. 00CH37121)}},\ }\href@noop {} {\ \textbf
  {\bibinfo {volume} {1}},\ \bibinfo {pages} {869} (\bibinfo {year}
  {2000})}\BibitemShut {NoStop}%
\bibitem [{\citenamefont {Megrant}\ \emph {et~al.}(2012)\citenamefont
  {Megrant}, \citenamefont {Neill}, \citenamefont {Barends}, \citenamefont
  {Chiaro}, \citenamefont {Chen}, \citenamefont {Feigl}, \citenamefont {Kelly},
  \citenamefont {Lucero}, \citenamefont {Mariantoni}, \citenamefont
  {O’Malley}, \citenamefont {Sank}, \citenamefont {Vainsencher},
  \citenamefont {Wenner}, \citenamefont {White}, \citenamefont {Yin},
  \citenamefont {Zhao}, \citenamefont {Palmstrøm}, \citenamefont {Martinis},\
  and\ \citenamefont {Cleland}}]{megrant2012}%
  \BibitemOpen
  \bibfield  {author} {\bibinfo {author} {\bibfnamefont {A.}~\bibnamefont
  {Megrant}}, \bibinfo {author} {\bibfnamefont {C.}~\bibnamefont {Neill}},
  \bibinfo {author} {\bibfnamefont {R.}~\bibnamefont {Barends}}, \bibinfo
  {author} {\bibfnamefont {B.}~\bibnamefont {Chiaro}}, \bibinfo {author}
  {\bibfnamefont {Y.}~\bibnamefont {Chen}}, \bibinfo {author} {\bibfnamefont
  {L.}~\bibnamefont {Feigl}}, \bibinfo {author} {\bibfnamefont
  {J.}~\bibnamefont {Kelly}}, \bibinfo {author} {\bibfnamefont
  {E.}~\bibnamefont {Lucero}}, \bibinfo {author} {\bibfnamefont
  {M.}~\bibnamefont {Mariantoni}}, \bibinfo {author} {\bibfnamefont {P.~J.~J.}\
  \bibnamefont {O’Malley}}, \bibinfo {author} {\bibfnamefont
  {D.}~\bibnamefont {Sank}}, \bibinfo {author} {\bibfnamefont {A.}~\bibnamefont
  {Vainsencher}}, \bibinfo {author} {\bibfnamefont {J.}~\bibnamefont {Wenner}},
  \bibinfo {author} {\bibfnamefont {T.~C.}\ \bibnamefont {White}}, \bibinfo
  {author} {\bibfnamefont {Y.}~\bibnamefont {Yin}}, \bibinfo {author}
  {\bibfnamefont {J.}~\bibnamefont {Zhao}}, \bibinfo {author} {\bibfnamefont
  {C.~J.}\ \bibnamefont {Palmstrøm}}, \bibinfo {author} {\bibfnamefont
  {J.~M.}\ \bibnamefont {Martinis}}, \ and\ \bibinfo {author} {\bibfnamefont
  {A.~N.}\ \bibnamefont {Cleland}},\ }\href {\doibase 10.1063/1.3693409}
  {\bibfield  {journal} {\bibinfo  {journal} {Applied Physics Letters}\
  }\textbf {\bibinfo {volume} {100}},\ \bibinfo {pages} {113510} (\bibinfo
  {year} {2012})}\BibitemShut {NoStop}%
\bibitem [{\citenamefont {Bienfait}\ \emph {et~al.}(2020)\citenamefont
  {Bienfait}, \citenamefont {Zhong}, \citenamefont {Chang}, \citenamefont
  {Chou}, \citenamefont {Conner}, \citenamefont {Dumur}, \citenamefont
  {Grebel}, \citenamefont {Peairs}, \citenamefont {Povey}, \citenamefont
  {Satzinger},\ and\ \citenamefont {Cleland}}]{bienfait2020}%
  \BibitemOpen
  \bibfield  {author} {\bibinfo {author} {\bibfnamefont {A.}~\bibnamefont
  {Bienfait}}, \bibinfo {author} {\bibfnamefont {Y.~P.}\ \bibnamefont {Zhong}},
  \bibinfo {author} {\bibfnamefont {H.-S.}\ \bibnamefont {Chang}}, \bibinfo
  {author} {\bibfnamefont {M.-H.}\ \bibnamefont {Chou}}, \bibinfo {author}
  {\bibfnamefont {C.~R.}\ \bibnamefont {Conner}}, \bibinfo {author}
  {\bibfnamefont {{\'E}.}~\bibnamefont {Dumur}}, \bibinfo {author}
  {\bibfnamefont {J.}~\bibnamefont {Grebel}}, \bibinfo {author} {\bibfnamefont
  {G.~A.}\ \bibnamefont {Peairs}}, \bibinfo {author} {\bibfnamefont {R.~G.}\
  \bibnamefont {Povey}}, \bibinfo {author} {\bibfnamefont {K.~J.}\ \bibnamefont
  {Satzinger}}, \ and\ \bibinfo {author} {\bibfnamefont {A.~N.}\ \bibnamefont
  {Cleland}},\ }\href {\doibase 10.1103/PhysRevX.10.021055} {\bibfield
  {journal} {\bibinfo  {journal} {Phys. Rev. X}\ }\textbf {\bibinfo {volume}
  {10}},\ \bibinfo {pages} {021055} (\bibinfo {year} {2020})}\BibitemShut
  {NoStop}%
\bibitem [{\citenamefont {Koch}\ \emph {et~al.}(2007)\citenamefont {Koch},
  \citenamefont {Yu}, \citenamefont {Gambetta}, \citenamefont {Houck},
  \citenamefont {Schuster}, \citenamefont {Majer}, \citenamefont {Blais},
  \citenamefont {Devoret}, \citenamefont {Girvin},\ and\ \citenamefont
  {Schoelkopf}}]{koch2007}%
  \BibitemOpen
  \bibfield  {author} {\bibinfo {author} {\bibfnamefont {J.}~\bibnamefont
  {Koch}}, \bibinfo {author} {\bibfnamefont {T.~M.}\ \bibnamefont {Yu}},
  \bibinfo {author} {\bibfnamefont {J.}~\bibnamefont {Gambetta}}, \bibinfo
  {author} {\bibfnamefont {A.~A.}\ \bibnamefont {Houck}}, \bibinfo {author}
  {\bibfnamefont {D.~I.}\ \bibnamefont {Schuster}}, \bibinfo {author}
  {\bibfnamefont {J.}~\bibnamefont {Majer}}, \bibinfo {author} {\bibfnamefont
  {A.}~\bibnamefont {Blais}}, \bibinfo {author} {\bibfnamefont {M.~H.}\
  \bibnamefont {Devoret}}, \bibinfo {author} {\bibfnamefont {S.~M.}\
  \bibnamefont {Girvin}}, \ and\ \bibinfo {author} {\bibfnamefont {R.~J.}\
  \bibnamefont {Schoelkopf}},\ }\href {\doibase 10.1103/PhysRevA.76.042319}
  {\bibfield  {journal} {\bibinfo  {journal} {Physical Review A}\ }\textbf
  {\bibinfo {volume} {76}},\ \bibinfo {pages} {042319} (\bibinfo {year}
  {2007})}\BibitemShut {NoStop}%
\bibitem [{\citenamefont {Barends}\ \emph {et~al.}(2013)\citenamefont
  {Barends}, \citenamefont {Kelly}, \citenamefont {Megrant}, \citenamefont
  {Sank}, \citenamefont {Jeffrey}, \citenamefont {Chen}, \citenamefont {Yin},
  \citenamefont {Chiaro}, \citenamefont {Mutus}, \citenamefont {Neill},
  \citenamefont {O'Malley}, \citenamefont {Roushan}, \citenamefont {Wenner},
  \citenamefont {White}, \citenamefont {Cleland},\ and\ \citenamefont
  {Martinis}}]{barends2013}%
  \BibitemOpen
  \bibfield  {author} {\bibinfo {author} {\bibfnamefont {R.}~\bibnamefont
  {Barends}}, \bibinfo {author} {\bibfnamefont {J.}~\bibnamefont {Kelly}},
  \bibinfo {author} {\bibfnamefont {A.}~\bibnamefont {Megrant}}, \bibinfo
  {author} {\bibfnamefont {D.}~\bibnamefont {Sank}}, \bibinfo {author}
  {\bibfnamefont {E.}~\bibnamefont {Jeffrey}}, \bibinfo {author} {\bibfnamefont
  {Y.}~\bibnamefont {Chen}}, \bibinfo {author} {\bibfnamefont {Y.}~\bibnamefont
  {Yin}}, \bibinfo {author} {\bibfnamefont {B.}~\bibnamefont {Chiaro}},
  \bibinfo {author} {\bibfnamefont {J.}~\bibnamefont {Mutus}}, \bibinfo
  {author} {\bibfnamefont {C.}~\bibnamefont {Neill}}, \bibinfo {author}
  {\bibfnamefont {P.}~\bibnamefont {O'Malley}}, \bibinfo {author}
  {\bibfnamefont {P.}~\bibnamefont {Roushan}}, \bibinfo {author} {\bibfnamefont
  {J.}~\bibnamefont {Wenner}}, \bibinfo {author} {\bibfnamefont {T.~C.}\
  \bibnamefont {White}}, \bibinfo {author} {\bibfnamefont {A.~N.}\ \bibnamefont
  {Cleland}}, \ and\ \bibinfo {author} {\bibfnamefont {J.~M.}\ \bibnamefont
  {Martinis}},\ }\href {\doibase 10.1103/PhysRevLett.111.080502} {\bibfield
  {journal} {\bibinfo  {journal} {Physical Review Letters}\ }\textbf {\bibinfo
  {volume} {111}},\ \bibinfo {pages} {080502} (\bibinfo {year}
  {2013})}\BibitemShut {NoStop}%
\end{thebibliography}%
\end{document}